\documentclass[aps,pre,twocolumn,superscriptaddress,showpacs]{revtex4}
\usepackage{amsfonts,amssymb,amsmath,latexsym,epsfig,graphicx} 
\begin{document}  

\title{Stickiness in Hamiltonian systems: from sharply divided to hierarchical phase space}

\author{Eduardo G. Altmann}
\email{edugalt@pks.mpg.de}
\affiliation{Max Planck Institute for the Physics of Complex Systems, N\"othnitzer Strasse 38, 01187 Dresden, Germany}

\author{Adilson E. Motter}
%\affiliation{Center for Nonlinear Studies and Complex Systems Group, Theoretical Division, 
\affiliation{CNLS and Theoretical Division,  Los Alamos National Laboratory, Los Alamos, NM 87545, USA}
\affiliation{Department of Physics and Astronomy, Northwestern University, Evanston, IL 60208, USA}

\author{Holger Kantz}
\affiliation{Max Planck Institute for the Physics of Complex Systems, N\"othnitzer Strasse 38, 01187 Dresden, Germany}

\date{\today}

\begin{abstract}
We investigate the dynamics of chaotic trajectories in simple yet 
physically important Hamiltonian systems with non-hierarchical borders
between regular and chaotic regions with positive measures. We show that the stickiness to the
border of the regular regions in
systems with such a sharply divided phase space occurs through one-parameter
families of marginally unstable periodic orbits and is characterized by an
exponent $\gamma=2$ for the asymptotic power-law decay of the distribution of
recurrence times.  Generic perturbations lead to systems with hierarchical
phase space, where the stickiness is apparently enhanced due to the presence of
infinitely many regular islands and Cantori. In this case, we show that the
distribution of recurrence times can be composed of a sum of exponentials
or a sum of power-laws, depending on the relative contribution of the primary
and secondary structures of the hierarchy. Numerical verification of our main
results are provided for area-preserving maps, mushroom billiards, and
the newly defined {\it magnetic} mushroom billiards.
\end{abstract}
\maketitle

\section{\label{sec.I} Introduction}

Hamiltonian systems usually exhibit divided phase space,
where regular and chaotic regions coexist. An important
property of chaotic trajectories in divided phase spaces is
the intermittent behavior with sporadically long periods of
time spent near the border of regular regions \cite{karney}.
Because of this stickiness and the ergodicity of the chaotic
regions, even small islands can have a large effect on 
global properties of the system, such as  transport
\cite{zaslavsky} and decay of correlations \cite{karney}. 
The stickiness can be quantified in terms of the distribution $P(T)$
of recurrence times $T$ of a typical trajectory to a pre-defined
recurrence region, usually taken away from regular islands.
For fully chaotic hyperbolic systems, the recurrence
time distribution (RTD) decays exponentially \cite{altmann}, while for Hamiltonian systems
with divided phase space the RTD has been argued to decay
approximately as a power law
$P(T) \sim T^{-\gamma'}$ for large $T$, where $\gamma'$ is a scaling exponent
\cite{karney,zaslavsky,meiss.ott,chirikov,ketzmerick}. For power-law decay, the cumulative
RTD is given by
\begin{equation}\label{eq.powerlaw}
Q(\tau) \equiv \sum_{T=\tau}^{\infty} P(T) \sim \tau^{-\gamma},
\end{equation}
where $\gamma=\gamma'-1$.  
 We say that a system has the property of
stickiness if $Q(\tau)$ decays at least as slowly as $\tau^{-\gamma}$
for some $\gamma>0$. The existence of a finite mean recurrence  time
implies $\gamma>1$ \cite{zaslavsky}.
Experimental evidence of stickiness has been observed, for
example,  in the transport of particles advected by fluid flows 
\cite{swinney} and in the fluctuations of conductance in chaotic 
cavities \cite{ketzmerick2}.

In Hamiltonian systems, the border between a regular and
chaotic region often presents a complex hierarchical structure
of Kolmogorov-Arnold-Moser (KAM) islands and Cantori. Cantori
are invariant Cantor sets that work as partial barriers to
the transport close to KAM islands \cite{mackay,meiss.ott}. 
Although many properties of this structure are well understood,
their consequences to the dynamics are still a matter
of intense study \cite{chirikov,kantz2,zaslavsky,ketzmerick}.
Even in the simplest case of two-dimensional systems, a number
of non-equivalent models have been proposed to describe the
stickiness of chaotic trajectories. Meiss and Ott introduced a
Markov-tree model that accounts for the hierarchical structure
and predicts a scaling exponent $\gamma=1.96$ \cite{meiss.ott}.
Chirikov and Shepelyansky used renormalization arguments at the breakdown of
the golden mean torus to predict a universal exponent~$\gamma=3$~\cite{chirikov}.
Zaslavsky and co-workers applied different renormalization arguments to 
the case of self-similar island chains, obtaining simple relations between~$\gamma$ and
the scaling properties of these chains~\cite{zaslavsky}.
There is also strong evidence of other stickiness mechanisms
in generic Hamiltonian systems \cite{zas.trap,rom-kedar,nontwist,dana2}.
The effects described in these previous works typically coexist and 
are responsible for finite-time numerical estimates of $\gamma$ lying in the
interval $1.5 \leq \gamma \leq 2.5$~\cite{ketzmerick}. However, because the convergence in Hamiltonian systems can take an
arbitrarily long time, in general it is not even clear whether the RTD
approximates a power-law distribution in the asymptotic limit. This slow convergence 
has inspired Motter and co-workers to introduce a new model that accounts
for the effects of the Cantori structure at finite times \cite{motter}.
While the general asymptotic behavior remains unresolved, the insight
provided by the study of classes of comprehensible Hamiltonian systems
is of fundamental importance.

In this paper, we investigate a mechanism for the stickiness of chaotic
trajectories in Hamiltonian systems with non-hierarchical borders between
regular and chaotic regions when both regions can have positive measure. 
Examples of systems with such a sharply divided phase space include 
piecewise-linear area-preserving maps \cite{lee,malovrh} and mushroom 
billiards \cite{nosso.mushroom,bunimovich.mushroom}.
In Hamiltonian systems, it is a common sense statement to relate the stickiness 
to the presence of hierarchical structures of KAM islands and Cantori. While 
this statement by itself is not wrong,  here we show that regular islands with 
non-hierarchical borders also stick.  Our results are valid for both zero- and
positive-measure regular islands, and include as particular cases previous findings 
for systems not exhibiting KAM islands, such as the Sinai and stadium 
billiards~\cite{primeiros,vivaldi,gaspard,armstead}. We find that the stickiness 
near non-hierarchical borders occurs due to the presence
of one-parameter families of marginally unstable periodic orbits (MUPOs).
%The simplest example of such MUPOs are the bouncing ball orbits present in billiards
%with parallel walls.
Based on the analysis of MUPOs, we show that the recurrence time does follow
a power-law distribution and that the scaling exponent is $\gamma=2$ in
two-dimensional sharply divided phase spaces, irrespective of other details of
the system. We also study the properties of generic perturbations of these systems.
Based on numerical simulations of mushroom billiards perturbed by magnetic fields, we observe that the
perturbations generate hierarchical structures of KAM islands and Cantori
of the same nature of those observed in typical Hamiltonian systems. The
perturbation of Hamiltonian systems with sharply divided phase space thus represents a new
route to Hamiltonian systems with hierarchical phase space. Previously
considered routes start either from fully integrable or fully chaotic
configurations. The onset of hierarchical structures introduces oscillations
in the RTD, which we show to be related to the
relative contribution of the primary and secondary structures of the hierarchy.

The paper is organized as follows.
In Sec.~\ref{sec.II}, we analyze the stickiness in sharply divided phase spaces.
In Sec.~\ref{sec.III}, we consider the effect of the hierarchical structures when
a system with sharply divided phase space is perturbed.
Discussion and conclusions are presented in the last section.

\section{\label{sec.II} Sharply divided phase space}

We study the stickiness of chaotic trajectories in two-dimensional
Hamiltonian systems with non-hierarchical borders between the regions of
chaotic and regular motion. As examples, we consider piecewise-linear 
area-preserving maps and mushroom billiards in Secs.~\ref{ssec.csm} and
\ref{ssec.parallel}, respectively. In contrast to the previously considered
stadium and Sinai billiards, in these systems both the chaotic 
and regular regions of the phase space have a positive measure. 
The scaling exponent $\gamma=2$ is derived in Sec.~\ref{ssec.exponent}.

\subsection{Piecewise-linear maps}\label{ssec.csm}

Consider two-dimensional area-preserving maps of the form
\begin{equation}\label{eq.map}
\begin{array}{ll}
y_{n+1} =  y_n+K f(x_n)\; & \mod~1,\\
x_{n+1} = x_n+y_{n+1}\;\; & \mod~1,
\end{array} 
\end{equation}
where~$K$ is a parameter that controls the nonlinearity.
For $f(x_n)=\sin(2\pi x_n)$, Eq.~(\ref{eq.map}) corresponds
to the standard map, which has served as a prototype
of Hamiltonian system in numerous studies of stickiness in
hierarchical phase space \cite{karney,zaslavsky,white,chirikov,ketzmerick}.

However, for $f(x_n)$ defined as a piecewise-linear function
of the interval~$x_n\in[0,1]$, the phase space of map (\ref{eq.map})
can be sharply divided in the sense that regular and chaotic
regions are separated by a simple curve~\cite{wojtkowski}. 
As shown in Refs.~\cite{wojtkowski,ashwin,adler}, the form and 
distribution of the regular regions in general depend on the function
$f$ and on the parameter $K$. In the case of hierarchical distribution
of islands, it has been shown that the stickiness of chaotic trajectories
leads to anomalous diffusion in the extended phase space of these maps 
\cite{lowenstein}.

\begin{figure}[!ht]
\centerline{
\includegraphics[width=\columnwidth]{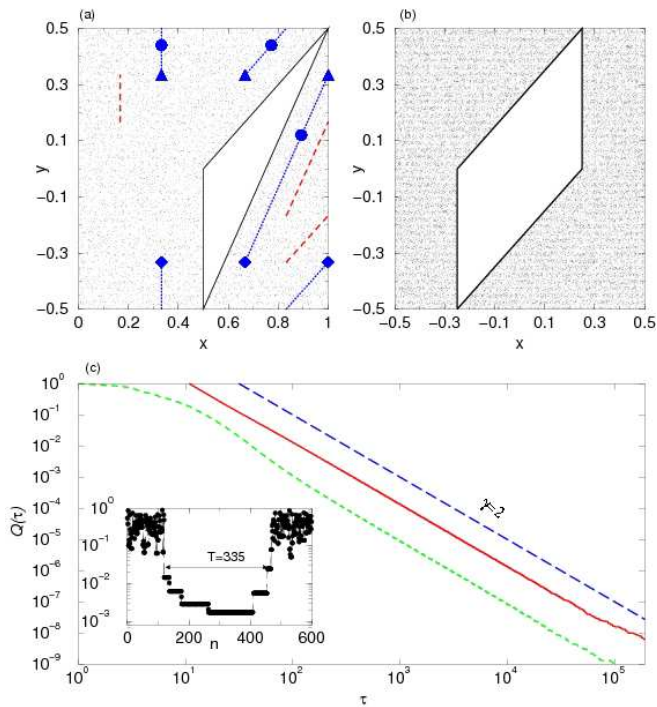}}
\caption{(Color Online) (a) Phase-space portrait of the  continuous sawtooth map
for $K=3/2$.
The dots correspond to $10^4$ iterations of a chaotic trajectory
and the blank region corresponds to the regular island. The straight lines
represent three different families of period-three MUPOs (see text),
and the different symbols correspond to specific MUPOs in one of these families. 
(b) Phase-space portrait of map~(\ref{eq.map})-(\ref{eq.lee}) for~$K=2$. 
We plot $-0.5 \le y \le 0.5$ in (a) and $-0.5 \leq x,y \leq 0.5$ in (b) for
visualization convenience.
(c) From bottom to top, the cumulative RTDs of the maps considered in (a) and (b) (multiplied by a factor $10$ for clarity). The upper curve is a
straight line with slope $\gamma=2$. 
Inset: distance of a chaotic trajectory to the the border of the island in (a)
during an event with recurrence time~$T=335$.
}
\label{fig.csmef}
\end{figure}

To quantify the stickiness in the case of sharp border, we consider 
two simple examples with a single regular island:

\noindent {\em (i)} The first example is obtained for
\begin{equation}\label{eq.csm}
f(x_n)=1-|2 x_n-1|,
\end{equation}
and was called {\em continuous sawtooth map} in Ref.~\cite{malovrh}. 
The phase space of this map is shown in Fig.~\ref{fig.csmef}(a) for  $K=
3/2$. It was argued in Ref.~\cite{malovrh} that in this case a single regular
island exists [Fig.~\ref{fig.csmef}(a), triangular region].  
As we show, the absence of other islands and the ergodicity of the
chaotic region do not rule out the possible existence of zero measure
sets of MUPOs in the chaotic region. In this paper we use the acronym
MUPOs to refer to periodic orbits in contact with the
chaotic component that have zero Lyapunov exponents and real eigenvalues
with modulus one. In sharply divided phase spaces,  we regard the borders
of regular islands as families of MUPOs whenever they are periodic. For example, for the continuous
sawtooth map, the following sets and their images correspond
to one-parameter families of period-three MUPOs: $\{x=1/6,1/6\le y\le 1/3\}$, 
$\{x=1/3, 1/3\le y\le 2/3\}$, and  $\{x=1/2, 1/2\le y\le 1\}$, where
the latter is at the border of the island. These families of MUPOs
correspond to the straight lines in Fig.~\ref{fig.csmef}(a). 

\noindent {\em (ii)} A second example of sharply divided phase space is obtained for
\begin{equation}\label{eq.lee}
f(x_n)=\left\{\begin{array}{lll}
     -x_n      &\mbox{if }\;\;\;\; 0 \leq  x_n  < 1/4, \\
     -1/2+x_n  &\mbox{if } 1/4 \leq  x_n  < 3/4, \\
    1-x_n      &\mbox{if } 3/4 \leq  x_n  \leq 1,
\end{array}
\right.
\end{equation}
as considered in Ref.~\cite{lee}. The properties of map
(\ref{eq.map})-(\ref{eq.lee}) with~$K=2$ are essentially the same of the
continuous sawtooth map  with~$K=3/2$~[Fig.~\ref{fig.csmef}(b) vs. Fig.~\ref{fig.csmef}(a)]. 
In particular, the phase space of map (\ref{eq.map})-(\ref{eq.lee}) has a single regular
island. The only relevant difference is that in this case there
is no other family of MUPOs apart from the border of the island. 

\begin{figure}[!ht]
\centerline{
\includegraphics[width=\columnwidth]{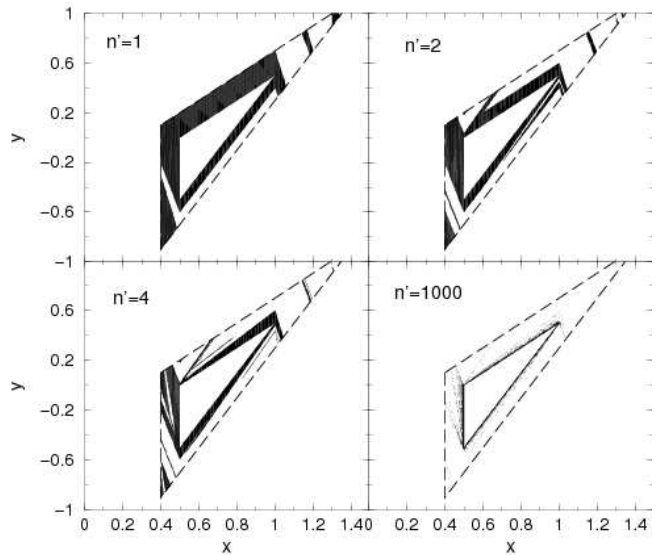}}
\caption{Phase-space portrait of the continuous sawtooth map
for $K=3/2$ showing the initial conditions of the trajectories
that remain inside the dashed triangle for at least $n'=1$, $2$, $4$ and $1000$
iterations of the map, respectively.
The inner triangle corresponds to the regular island. We
plot~$0 \leq x \leq 1.5$ and $-1\leq x \leq1$ for visualization convenience.} 
\label{fig.csmremains}
\end{figure}

In the stickiness of a chaotic trajectory, the trajectory approaches a
family of MUPOs and follows a nearly periodic dynamics for a long period
of time before leaving the neighborhood of the MUPOs [inset of Fig.~\ref{fig.csmef}(c)]. 
In Fig.~\ref{fig.csmremains}, we show the trajectories that stick to the 
MUPOs at the border of the regular island and remain in the
neighborhood of the island after $n'=1$, $2$, $4$ and $1000$ iterations
of the continuous sawtooth map. In general, the closer to the island the longer it will take
for the trajectory to leave. This stickiness is properly quantified in terms of the
RTD for a recurrence region taken apart from islands. 
We have performed numerical simulations for two different configurations
presenting sharply divided phase space:
the continuous sawtooth map with~$K=3/2$ [Fig.~\ref{fig.csmef}(a)] and
map~(\ref{eq.map})-(\ref{eq.lee}) with~$K=2$
[Fig.~\ref{fig.csmef}(b)]. As shown in Fig.~\ref{fig.csmef}(c), in both cases
the cumulative RTD is best approximated by a power law with scaling exponent
$\gamma=2$.
 
We have found similar results for other piecewise-linear
area-preserving maps with polygonal islands, which were obtained 
from the maps considered above for different choices of the
parameter~$K$ and from a different map studied in Ref.~\cite{wojtkowski}.
There are cases when the results described in this section
and the theory of Sec.~\ref{ssec.exponent} do not apply, such as when the
regular islands are elliptical and the outermost torus is 
quasi-periodic. Nevertheless, we observed that in many 
of these cases the exponent~$\gamma=2$ also fits the power-law tails
of the numerically obtained RTD.

%%%%%%%%%%%%%%%%%%%%%%%%%%%%%%%%%%%%%%%%%%%%%%%%%%%%%%%%%%%%%%%%%%%%%%%%%%%%%%%%%%%%%%%%%%%%%%%%%%%%%%%%%%%%%%%%%
   
\subsection{Mushroom billiards}\label{ssec.parallel}
  
\begin{figure}[!ht]
\centerline{
\includegraphics[width=\columnwidth]{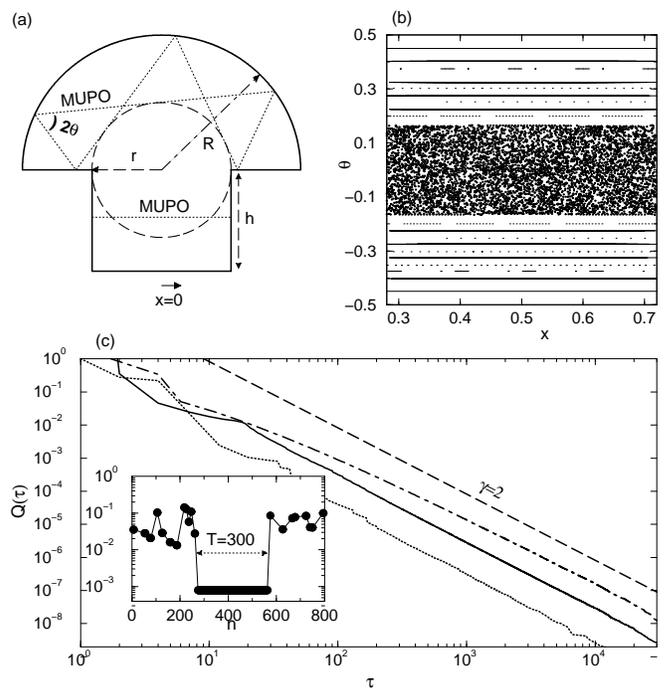}}
\caption{(a) Typical mushroom billiard, defined by the geometric parameters
  $(r,R,h)$. Two MUPOs are shown in dotted lines. (b) Phase-space
  representation of the semi-circular hat of the mushroom billiard
  with~$r/R=0.5$. (c) From bottom to top, RTDs for 
  $r/R=0.6$, $0.75$, and $0.5$ (multiplied by a factor $2$ for clarity). The upper curve is a straight line with
  slope~$\gamma=2$. Inset: distance of a chaotic trajectory to the border of the island during
  an event with recurrence time~$T=300$. }
\label{fig.mushroom}
\end{figure}

Billiards can be used as simple models in the study of  Hamiltonian systems.
Recently, Bunimovich introduced the so called mushroom billiards \cite{bunimovich.mushroom},
which are billiards that have a single regular and a single chaotic ergodic region.
A typical mushroom billiard is defined by a semi-circle (hat) placed on top of a rectangle (foot),
as depicted in Fig.~\ref{fig.mushroom}(a). The phase space is described by the
normalized position~$x$ on the boundary of the billiard and
angle~$\theta\in[-0.5,0.5]$ with respect to the normal vector right after the
specular reflection. 
%The time is counted discreetly as the number of collisions. 
The regular region corresponds to the orbits in 
the hat of the mushroom that never cross the dashed circle of radius~$r$ in
Fig.~\ref{fig.mushroom}(a). 
The border between the regular and chaotic region of the
mushroom billiard is therefore non-hierarchical,
%and the phase space is thus sharply divided, 
as shown in fig~\ref{fig.mushroom}(b).

Mushroom billiards  have two different classes
of MUPOs, as illustrated in Fig.~\ref{fig.mushroom}(a). One of them corresponds to orbits bouncing between the parallel walls in
the foot of the mushroom. Similar MUPOs are also found in many other billiards with parallel
walls, such as the Sinai and stadium billiards~\cite{primeiros,vivaldi,gaspard,armstead}. The other and more interesting class
of MUPOs corresponds to periodic orbits in the chaotic region that never leave the hat of
the mushroom. In a previous study \cite{nosso.mushroom}, we have shown that there is usually a 
complex distribution of these MUPOs close to the regular region and in contact
with the chaotic component. The border of the regular island 
can be regarded as MUPOs of this class.  We have found similar MUPOs in other billiards with circular component,
such as annular billiards~\cite{annular}.

A relevant point concerning the stickiness in mushroom billiards is that 
the scaling exponent of the cumulative RTD is again $\gamma=2$,
regardless of the control parameter~$0<r/R \leq 1$, as shown in
Fig.~\ref{fig.mushroom}(c). In this case, the whole foot of the mushroom is taken as the
recurrence region in order to avoid the trivial parallel wall MUPOs.
The injection and escape mechanism of the
chaotic trajectories near the island are slightly different from those
observed in the continuous sawtooth map because in mushroom billiards there are escaping regions
tangent to the island and the injection and escape occur in a single
iteration [inset of Fig.~\ref{fig.csmef}(c) vs. inset of Fig.~\ref{fig.mushroom}(c)] \cite{nosso.mushroom}. 
However, from a more fundamental point of view, the stickiness is remarkably
similar in these systems because in both cases the stickiness is mediated by
MUPOs and the RTD has an exponent~$\gamma=2$. 
Altogether, this suggests the possible existence of a universal scenario
for the stickiness of chaotic trajectories in systems with divided phase
space, as considered in the next section.

\subsection{Scaling exponent: Theory\label{ssec.exponent}}

We now derive the scaling exponent $\gamma=2$ for the cumulative 
RTD of two-dimensional systems with sharply divided phase space. 
Our theory applies to the class of systems presenting
one-dimensional families of MUPOs. This includes as particular cases
some systems without regular islands, such as the stadium and 
Sinai billiards, whose cumulative RTDs are known to be governed 
by an exponent $\gamma=2$~\cite{gaspard,armstead}. Most 
importantly, our results also apply to systems with mixed phase
space, such as those considered in the previous sections. The theory remains
valid when the orbit at border of the regular region is quasi-periodic and
the first-escape region is tangent to the border. That is, if there are
trajectories arbitrarily  close to the island that move away in one or few
time steps, such as in the mushroom billiards \cite{nosso.mushroom}. 

\begin{figure}[!ht]
\centerline{
\includegraphics[width=\columnwidth]{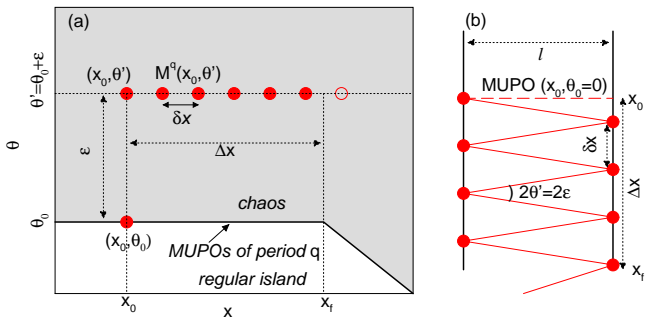}}
\caption{(a) Illustration of the dynamics of a perturbed MUPO~($x_0,\theta_0 +\varepsilon$)
  in the phase space (see text). (b) The corresponding dynamics in the configuration
  space of a billiard with parallel walls ($q=2$).}\label{fig.scaling}
\end{figure}

The essential features of the systems considered in the previous sections
are captured by an area-preserving map $M(x,\theta)$ defined on the torus
and that contains a one-parameter family of MUPOs of period~$q$.
For concreteness, we assume that the family of MUPOs is $\{x_i\le x\le x_f$, $\theta=\theta_0\}$.
The phase space of map~$M(x,\theta)$ is sketched in Fig.~\ref{fig.scaling}(a) and a possible
configuration space in Fig.~\ref{fig.scaling}(b). The following analysis does not
depend on whether the MUPOs are in the chaotic sea or at the border of a regular
island.

Consider small perturbations of a MUPO~$(x_0,\theta_0)$:
\begin{itemize}
\item[({\em i})] If $(x',\theta')=(x_0+\epsilon_x,\theta_0)$ and $x_i\le x'\le x_f$,
  another periodic orbit of
  the set of MUPOs is obtained. In this
  case $M^q(x',\theta_0)=(x',\theta_0)$, that shows that the perturbation
  neither grows nor shrinks. 
\item[({\em ii})] If $(x',\theta')=(x_0,\theta_0+\varepsilon)$, the perturbation in the
  $\theta$ direction does not grow. On the
  other hand, in the~$x$ direction the trajectory is not strictly periodic anymore and
  there is a displacement~$\delta x$ every period~$q$:
  $M^q(x_0,\theta')=(x_0+\delta x,\theta')$.   
\end{itemize}

Both effects~{\it (i)} and~{\it (ii)} have to be taken into account when a
generic perturbation is considered:
\begin{equation}\label{eq.perturbation}
  M^q(x',\theta') \equiv M^q(x_0+\epsilon_x, \theta_0+\varepsilon)=(x'+\delta
  x,\theta').
\end{equation}
After $q$ iterations, the same arguments used above for~$(x',\theta')$ apply to~$(x'+\delta x,\theta')$.
We thus see that the perturbed trajectory follows the
dynamics~(\ref{eq.perturbation}), remaining at a constant
distance~$\varepsilon$ from the family of MUPOs, until it travels~$\Delta
x=x_f-x_0$ reaching the end $x=x_f$ of the family of MUPOs (see Fig.~\ref{fig.scaling})~\cite{xf}.
We note that Eq.~(\ref{eq.perturbation}) implies a linear growth of the perturbation in
time, what is consistent with the marginal instability of the fixed point that forbids
exponential growth of perturbations.

The displacement~$\delta x$ is related to the difference between the frequency of the
perturbed and unperturbed orbits, and can therefore be approximated linearly as
\begin{equation}~\label{eq.6}
\delta x = D \varepsilon,
\end{equation}
in the limit of small~$\varepsilon$.  For the continuous sawtooth
map with~$K=3/2$, one obtains~$D=6$. In the
case of billiards with parallel walls~$D = 2 l$, where~$l$ is the distance between the walls.
For MUPOs in circular like billiards, such as mushroom and annular
billiards,~$D=2q R$, where $R$ is the radius of the circle. 

The time a perturbed trajectory takes to reach~$x_f$ and escape from the
dynamics~(\ref{eq.perturbation}) is given by
\begin{equation}\label{eq.tepsilon}
T=\frac{\Delta x}{\delta x} \sim \frac{1}{\varepsilon} 
\end{equation}
for small~$\varepsilon$. In what follows, we see
that this time is equivalent to the recurrence time if the initial
conditions are chosen properly. Relation~(\ref{eq.tepsilon})
shows that the smaller the perturbation the longer the time the trajectory
takes to escape.  The asymptotic distribution of escape times~$P(T)$ as a
function of the distribution of perturbations~$p(\varepsilon)$ is given by
\begin{equation}\label{eq.gamma2}
P(T) = \frac{p(\varepsilon)}{|d T/d\varepsilon|} \sim p(\varepsilon) \varepsilon^2,
\; \text{with}\; \varepsilon\sim 1/T.
\end{equation}
The distribution~$p(\varepsilon)$ depends on the choice of the initial
conditions. 

For instance, choosing the initial conditions in the neighborhood of the family
of MUPOs leads to a rapid convergence of~$p(\varepsilon)$ to the invariant measure
of the system. In this case,  $p(\varepsilon)$ can be asymptotically regarded as a constant.
From
Eqs.~(\ref{eq.tepsilon}) and (\ref{eq.gamma2}), we obtain $\gamma'_{tr}=2$ for  the
power-law exponent of the distribution of escape times, or~$\gamma_{tr}=1$ for
the cumulative distribution. This description is not valid when recurrences are
calculated, because the initial
conditions are chosen inside the recurrence region and thus {\em away} from
the MUPOs. In this case, the convergence of~$p(\varepsilon)$ to the invariant measure
is much slower (algebraically for the stadium billiard~\cite{armstead}) and for any
finite time $p(\varepsilon)\rightarrow0$ for~$\varepsilon\rightarrow0$.
However, we show in the Appendix that the scaling exponent for this second case can
be derived from the first: the power-law exponent increases by~$+1$ when the initial conditions
are taken away from the MUPOs, i.e., $\gamma=\gamma_{tr}+1\;$ (see also Refs.~\cite{pikovsky,armstead,chirikov,meiss}).
In systems with a one-parameter family of MUPOs, this leads to the asymptotic exponent~$\gamma=2$
for the cumulative RTD, in agreement with our numerical results.

Since every family of MUPOs contributes with the same exponent~$\gamma=2$
asymptotically, the exponent does not depend on the possible presence of other
families of MUPOs in addition to the one at the border of the regular islands.
Indeed, a large number of other families is observed in mushroom 
billiards~\cite{nosso.mushroom}, a small number in the continuous sawtooth map
and none in the case of map~(\ref{eq.map})-(\ref{eq.lee}), and the scaling exponent
is the same in each of these cases.
  
It would be interesting to verify the generality of the
exponent~$\gamma=2$~\cite{inside}, for example, by investigating other
systems presenting sharply divided phase space~\cite{debievre}. The stickiness through MUPOs described
above resembles the mechanism underlying stickiness and anomalous diffusion in one-dimensional maps 
with marginally unstable fixed points~\cite{zumofen}. Here we have considered two-dimensional
systems and we believe that similar results hold true in higher dimensional
Hamiltonian systems. 

\section{\label{sec.III} Hierarchical phase space}

Now we consider perturbations to systems with sharply divided phase space. 
This leads us to the problem of stickiness in
Hamiltonian systems with the usual complex hierarchy of
infinitely many KAM islands and Cantori. As a model
system we consider the mushroom billiard perturbed by
a magnetic field, a system that we refer to as the
{\it magnetic} mushroom billiard  and that allows for a
direct comparison between the effects of hierarchical
and non-hierarchical borders.

\subsection{Perturbation of non-hierarchical borders}

A common feature of the systems considered in the previous
sections is that their dynamics is piecewise smooth and
presents abrupt changes. These abrupt changes,
generated by non-smooth functions $f$ in map (\ref{eq.map})
and sharp corners in the mushroom billiard,
are responsible for the creation of sharply divided
phase spaces. Generic perturbations of these systems are
expected to smooth down the dynamics and introduce
hierarchies of KAM islands and Cantori. 
Examples of such perturbations include to smoothen
functions (\ref{eq.csm}) or (\ref{eq.lee}) in the case of
piecewise-linear maps and soften the walls
in the case of mushroom billiards. In the case of a billiard with
charged particles, we can also perturb the system with a magnetic
field, as studied below.

Consider the mushroom billiard studied in Sec.~\ref{ssec.parallel}
subject to uniform transverse magnetic field~$B$ and consider the
dynamics of charged particles within this billiard.
Due to the Lorentz force, the charged particles move on
circular orbits. 
We choose the charge of the particles and orientation of the magnetic
field such that the trajectories are oriented counter-clockwise
and have radius
\begin{equation}
 L \propto \frac{1}{B}\;,
\end{equation}
which is used as a control parameter. This parameter has to be
compared with the geometric scales of the billiard defined in 
Fig.~\ref{fig.mushroom}(a) (in our simulations we use $R=2$ and $r=1$). 
The unperturbed mushroom billiard corresponds to $L=\infty$.

\begin{figure}[!ht]
\centerline{
\includegraphics[width=\columnwidth]{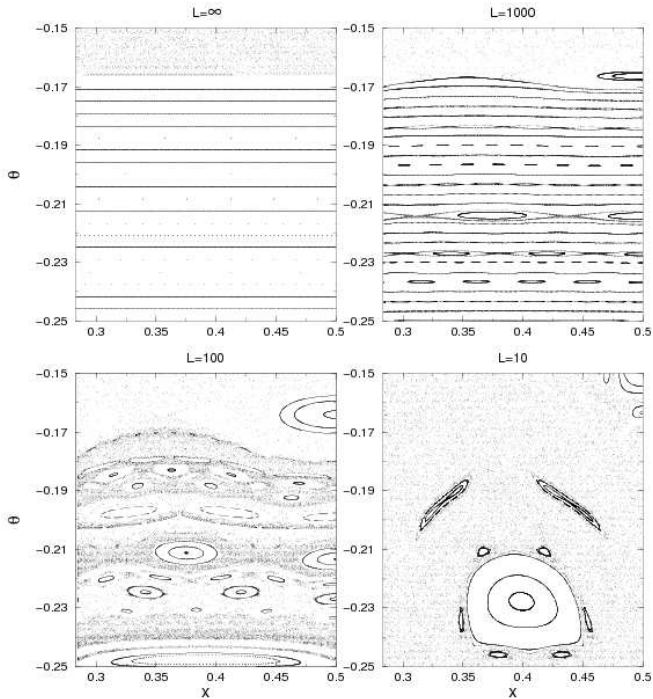}}
\caption{Magnification of the phase-space portrait of the magnetic
         mushroom billiard at the border between the chaotic
         and regular regions for $r/R=0.5$ and various values
	 of the magnetic field. }
\label{fig.Rb}
\end{figure}

Previous works on magnetic billiards~\cite{robnik} have shown that
the curvature of the trajectories often leads to the creation of
KAM  tori~\cite{breymann,gregorio} in  fully chaotic systems and
chaotic regions~\cite{meplan} in integrable systems.
Mushroom billiards have both integrable and chaotic regions in the
phase space and both effects are expected to take place.
More interestingly, mushroom billiards also have MUPOs that are
expected to undergo a transformation when the system is perturbed.
Indeed, because the eigenvalues associated to these orbits are
real and have modulus $1$, arbitrarily small perturbations are
expected to generate elliptic or saddle points in the neighborhood
of the regular island of the unperturbed billiard.
These effects of the magnetic field in the mushroom billiard
are shown in Fig.~\ref{fig.Rb}, where a representative magnification
of the phase space at the border of chaos is shown for different
values of~$L$.  The hierarchy of KAM islands and Cantori are clearly
visible, providing evidence that the complete picture of Hamiltonian
chaos is obtained in magnetic mushroom billiards.

 \begin{figure}[!ht]
\centerline{
\includegraphics[width=\columnwidth]{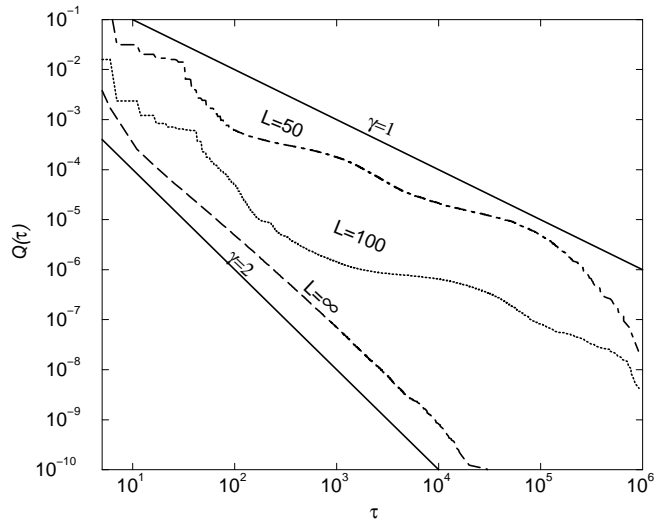}}
\caption{Cumulative RTDs for the magnetic mushroom billiard with~$r/R=0.5$ and
  different values of the magnetic field. From bottom to top the lines represent:
  a power law with $\gamma=2$, the numerical results for $L=\infty$ 
  (shifted downward by two decades for clarity), $L=100$ (shifted downward by one decade)
  and $L=50$, and a power law with~$\gamma=1$.}
\label{fig.cum}
\end{figure}

The emergence of complex structures of KAM islands in the phase space
influences the stickiness, as shown in Fig.~\ref{fig.cum} for  RTD of 
magnetic mushroom billiards with~$L=100$ and $L=50$. Comparing these
distributions with those of the unperturbed system~($L=\infty$), we note
the presence of fluctuations around a slower power-law tendency ($\gamma < 2$). 
This  result indicates that, as intuitively expected, a
hierarchical border sticks the trajectories in a more effective way than
a non-hierarchical border. While the presence of a single family of MUPOs in
a hierarchical phase space would be enough to guarantee $\gamma \leq 2$, we
observe that, generically, all the families of MUPOs disappear.
One could expect that the outermost torus of a regular island,
which is marginally unstable, could play the role of the MUPOs described
in Sec.~\ref{sec.II}. However, there is usually an infinite number of Cantori
that accumulate near the island invalidating relations~(\ref{eq.perturbation})
and (\ref{eq.6}) 
and thus the derivation of the exponent~$\gamma=2$. In the next section we study
carefully the effect of such a hierarchical border on the stickiness.

%%%%%%%%%%%%%%%%%%%%%%%%%%%%%%%%%%%%%%%%%%%%%%%%%%%%%%%%%%%%%%%%%%%%%%%%%%%%%%%%%%%%%%%%%%%%%%%%%%%%%%%%%%%%%%%%%

\subsection{Hierarchical phase-space scenarios}

We now investigate the origin of the oscillations
%and apparently slower decay of 
in the RTDs shown in Fig.~\ref{fig.cum}. 

\begin{figure}[!ht]
\centerline{
\includegraphics[width=\columnwidth]{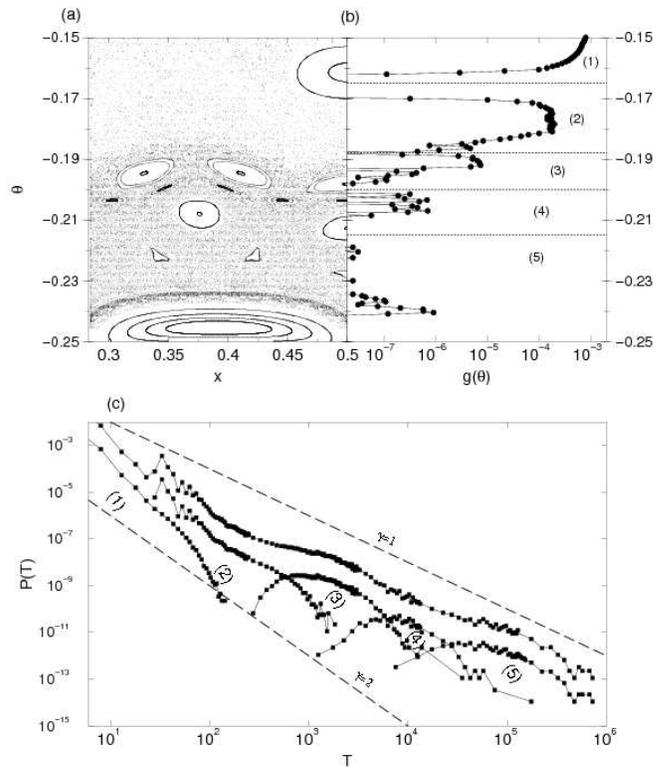}}
\caption{Analysis of the magnetic mushroom billiard with $L=50$:
  (a) phase-space magnification at the border of chaos; 
  (b) fraction~$g(\theta)$ of recurrences that have $\theta$ as their minimal angle;
  (c) the RTD of all trajectories (upper solid curve) and 
       the RTDs of the trajectories in regions~(1)-(5) of (b) (lower solid curves).
       The lower curves in (c) are divided by 10 for clarity.} 
\label{fig.rb50}
\end{figure}

We focus initially on the parameter~$L=50$. For this parameter, many KAM tori
are destroyed but the chain of islands and Cantori are still clearly visible
in the phase space, as shown in Fig.~\ref{fig.rb50}. 
The different density of points seen in Fig.~\ref{fig.rb50}(a) is related to the
presence of chains of islands and Cantori acting as partial barriers~\cite{mackay} to the transport
in the~$\theta$ direction.  In order to associate the presence of these
barriers to the RTD, we study the minimum distance between the trajectory and the
main island before the trajectory leaves the neighborhood of the island and
visits the recurrence region. In our simulations we use the minimum collision
angle~$\theta$ of the trajectory as a measure of the distance because the
barriers mimic the original tori and have approximately constant $\theta$,
and we take the foot of the mushroom billiard as the recurrence region.
The fraction of events that have a minimum angle~$\theta$ is defined as 
$g(\theta) d\theta=\eta_{\theta}/\eta$, where $\eta_{\theta}$ is the number
of recurrences that have a minimum angle in the interval [$\theta,\theta+d\theta$] 
and $\eta$ is the total number of  recurrences.  Numerical results for
$g(\theta)$ with $L=50$ are shown in Fig.~\ref{fig.rb50}(b). The function
$g(\theta)$ goes to zero at the angles that correspond to the position of
the barriers because the trajectories that manage to pass a barrier quickly
spread throughout the next chaotic region. From the behavior of $g(\theta)$ 
in Fig.~\ref{fig.rb50}(b), we can resolve $5$ different regions limited by
these barriers. To associate these regions with the RTD, we label all the
recurrence events from (1) to (5) according to the number of regions the
trajectory penetrates before returning to the recurrence region.
The RTD of each of these groups of recurrence events are shown in
Fig.~\ref{fig.rb50}(c). The RTD of all the events corresponds to the
sum of these partial RTDs and is shown in the same figure (upper solid curve).
The partial RTD of each region (1)-(5) presents a relatively peaked maximum
followed by an exponential decay.  Accordingly, most of the
orbits that have the same recurrence time~$T$ penetrate the same
number of barriers [note the logarithmic scale in Fig.~\ref{fig.rb50}(c)].
These results indicate that, for $T<10^6$, the stickiness is dominated
by the primary chain of barriers around the main regular island,
that is, the contribution of barriers associated to secondary islands
is negligible. These results also show that the oscillations observed in
the RTD around the power-law behavior are intrinsically associated to
the presence of the barriers in the phase space.

These stickiness properties agree well with the predictions
of the model proposed by Motter {\it et al.} in Ref.~\cite{motter}.
In that paper, the hierarchy of Cantori is modeled by a chain of coupled hyperbolic systems, where each hyperbolic system
models the area of the phase space limited by successive Cantori.
One of the strengths of this model is that it predicts not only
the asymptotic behavior of the non-hyperbolic dynamics around KAM
islands but also the finite-time dynamics assessable in numerical
simulations and experiments. The model predicts that the survival probability of particles in the neighborhood
of KAM islands fluctuates around a power law and is composed of a sum
of exponentials associated to the Cantori.
Our results in Fig.~\ref{fig.rb50} show that this behavior is indeed
present in real Hamiltonian systems. As shown below, this picture changes when
secondary structures of the hierarchy are relevant. This more general
stickiness scenario is observed in the mushroom billiard for larger
values of the magnetic field (e.g.,~$L=10$).  

\begin{figure}[!ht]
\centerline{
\includegraphics[width=\columnwidth]{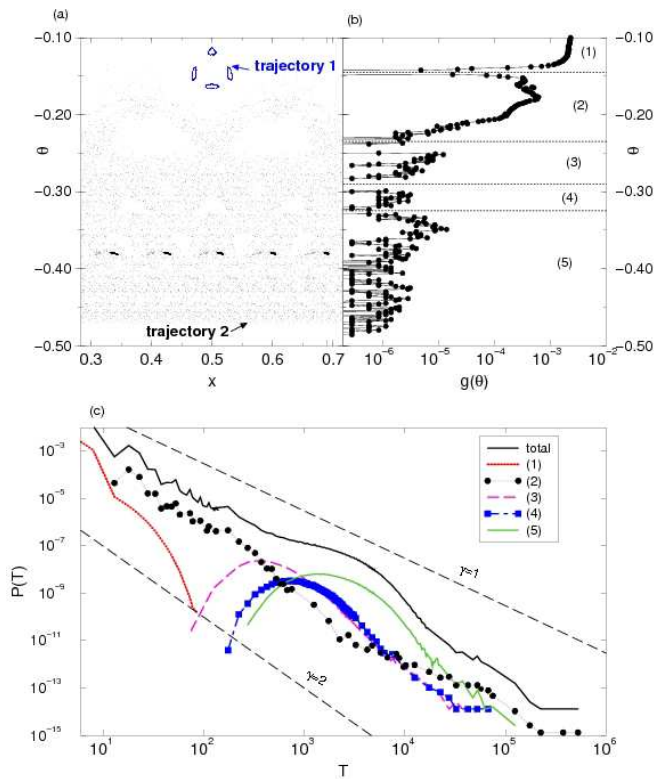}}
\caption{(Color Online) Analysis of the magnetic mushroom billiard with $L=10$.
   (a) Phase-space magnification with two typical sticking trajectories with recurrence time
  $T\approx 8\;10^4$: trajectory 1 sticks near the upper island and trajectory 2
  fills the chaotic region. (b) Fraction~$g(\theta)$ of recurrences that have $\theta$ as their minimal
  angle. (c) The RTD of all trajectories (upper solid curve) and 
  the RTDs of the trajectories in regions~(1)-(5) of (b) (see legend).
  The lower curves in (c) are divided by 10 for clarity.} 
\label{fig.rb10}
\end{figure}

In Fig.~\ref{fig.rb10}, we show the same as in Fig.~\ref{fig.rb50} for the
parameter~$L=10$.  The effect of the primary barriers is still important,
as shown in Fig.~\ref{fig.rb10}(b) where these barriers correspond to zeros
of $g(\theta)$. However, as shown in Fig.~\ref{fig.rb10}(c), the partial
RTDs corresponding to regions (1)-(5) exhibit a power-law rather
than an exponential decay. For instance, the RTD of trajectories
belonging to region (2) exhibits an approximate power-law  decay
that makes these recurrence events dominant
not only for small times ($10<T<500$) but also for very large times
($T\approx 10^5$).  On the other hand, the RTD of events
associated to region (4) does not dominate the (total) RTD at any time.
The slower decay of the RTD of region (2) is a
consequence of the stickiness to the chain of secondary islands shown at
the top of Fig.~\ref{fig.rb10}(a). In this figure, we show two representative
trajectories with recurrence time~$T\approx8\;10^4$. The first (trajectory 1)
penetrates only two regions and sticks to a secondary island.
The second (trajectory 2) penetrates five regions and approaches
the main island. 
%The presence of secondary islands at each level,
%which is expected to be common in typical Hamiltonian systems, shows the
%limitations of stickiness models based on a single
%chain of Cantori~\cite{ketzmerick}. 
In the context of stochastic
models~\cite{hanson,white},  asymptotic effects of secondary islands 
can be accounted for by the Markov-tree models~\cite{meiss.ott}. 
In a deterministic framework, the full hierarchy of islands can be accommodated within
a chain model of non-hyperbolic systems or a tree model of 
hyperbolic systems, which are straightforward generalizations of the
model introduced by Motter {\it et al.}~\cite{motter}.

%%%%%%%%%%%%%%%%%%%%%%%%%%%%%%%%%%%%%%%%%%%%%%%%%%%%%%%%%%%%%%%%%%%%%%%%%%%%%%%%%%%%%%%%%%%%%%%%%%%%%%%%%%%%%%%%%
%                             Conclusions
%%%%%%%%%%%%%%%%%%%%%%%%%%%%%%%%%%%%%%%%%%%%%%%%%%%%%%%%%%%%%%%%%%%%%%%%%%%%%%%%%%%%%%%%%%%%%%%%%%%%%%%%%%%%%%%%%

\section{\label{sec.IV} Conclusions}

We have studied the stickiness of chaotic trajectories in Hamiltonian systems
with sharply divided phase space, which are characterized by non-hierarchical
borders between the regions of chaotic and regular motion. The stickiness occurs
through the approach to one-parameter families of MUPOs in contact with the
chaotic region. The main characteristics of this stickiness scenario are the
exponent~$\gamma=2$ for the power-law decay of the cumulative RTD and the long
intervals of regular motion at a constant distance from families of MUPOs. 
Dynamical systems described by this scenario include mushroom billiards and 
various piecewise-linear area-preserving maps. 

Generic perturbations applied to systems with sharply divided phase space
destroy the MUPOs and introduce hierarchies of regular islands and Cantori.
We believe that these perturbations can serve as a new paradigm to the study
of stickiness in generic Hamiltonian systems. Using as an example
mushroom billiards perturbed by a transverse magnetic field, we characterize
two different scenarios of stickiness in the presence of perturbations.
For small perturbations, the stickiness is dominated by the primary
chain of Cantori, which work as partial barriers to the transport
around the main regular island. In this case, the RTD is composed of a sum
of exponential distributions associated to the probability of crossing each of these barriers.
%~\cite{hanson}. 
For increasing perturbations, the primary barriers weaken while the secondary
islands and the corresponding sticking regions grow. For large perturbations,
the stickiness of the secondary structures becomes relevant and the
exponential components of the RTD are converted themselves into
power-law distributions. This provides direct evidence of the effects of
Cantori structures at finite times, in strong support of the model
introduced in Ref.~\cite{motter} and its generalizations.

The asymptotic behavior of the RTD, which has been a matter of considerable
recent debate~\cite{chirikov,ketzmerick}, cannot be resolved alone by numerical experiments.
Our simulations suggest that the hierarchical structures enhance the stickiness
of non-hierarchical borders, what would lead to an upper bound $2$ for the scaling
exponent~$\gamma$.  This upper bound is guaranteed when the phase space
has one or more families of the MUPOs described in Sec.~\ref{sec.II}. However,
in general, this numerical evidence of upper bound should be taken with caution
because one cannot neglect the possibility that the hierarchical structures
will reduce the stickiness for asymptotically large times. For general Hamiltonian
systems, even the question of whether the oscillations in the RTD vanish
asymptotically, giving rise to a well defined power-law exponent,
is a problem yet to be settled. Our results provide an answer to
this question for an important class of Hamiltonian systems with
sharply divided phase space.

\begin{acknowledgments} 
%E.~G.~A. thanks
The authors thank G. Cristadoro, S. Denysov, and R. Klages for helpful discussions.
E.~G.~A. was supported by CAPES (Brazil) and DAAD (Germany).
A.~E.~M. was supported by the U.S. Department of Energy under Contract No.~W-7405-ENG-36.
\end{acknowledgments}

\section*{Appendix}

Consider a uniform distribution of initial conditions in the neighborhood of 
a sticking region of the phase space, as usually studied in problems of 
transient chaos, and consider the time it takes for the corresponding
trajectories to escape to a pre-defined region away from the sticking region. 
The distribution~$S(\tau)$ of escape times longer than~$\tau$ is proportional
to the measure $\mu(\tau)$ of the region of the phase space to which the
trajectories stick for a time longer than~$\tau$. Due to the ergodicity, 
\begin{equation}\label{eq.transient}
S(\tau) \propto \mu(\tau)=\frac{t_\tau}{t},
\end{equation}
where $t_\tau$ is the total time spent inside the sticking region and~$t$ is
the total observation time.

On the other hand, in the study of recurrence problems, one usually initializes a single
trajectory in a recurrence region {\em away} from the sticking region and computes
the time~$T$ the trajectory takes to return to that region. 
If the trajectory is followed for a long time $t$, the cumulative RTD is
\begin{equation}\label{eq.defptau}
Q(\tau) = \frac{N_\tau}{N},
\end{equation}
where~$N_\tau$ is the number of recurrences with time~$T\ge\tau$ and~$N$ is the total
number of recurrences observed in time~$t$.
The relation between the times in
Eq.~(\ref{eq.transient}) and the number of recurrences in
Eq.~(\ref{eq.defptau}) is given by~\cite{chirikov}
\begin{eqnarray}\label{eq.transf}
t       &\sim   N \; \langle T \rangle,\\
t_\tau  &\sim   N_\tau \; \tau,
\end{eqnarray}
where $\langle T \rangle$ is the average recurrence time.
Altogether, this leads to  
\begin{equation}\label{eq.qs}
Q(\tau) \sim \frac{S(\tau)}{\tau}\;.
\end{equation}
In particular, if the escape times follow a power-law distribution
$S(\tau)\sim \tau^{-\gamma_{tr}}$, then the cumulative RTD is
$Q(\tau) \sim \tau^{-\gamma}$ where
\begin{equation}
 \gamma=\gamma_{tr}+1\;.
\end{equation}
An equivalent relation was obtained in Ref.~\cite{pikovsky} for
chaotic scattering problems, another example where the trajectories are
initialized away from the sticking region.


\begin{references}
\bibitem{xf} If the MUPOs are at the border of a regular island, $x_f$
  represents either the vertex of a polygonal island [see
  Fig.~\ref{fig.csmef}(a,b)] or, as in the case of circular-like billiards, the
  point of tangency between the first escaping region and the island~(see Ref.~\cite{nosso.mushroom}).
\end{references}
\end{document}